\documentclass{article}
\usepackage[english]{babel}
\usepackage{float}
\usepackage{booktabs}
\usepackage{graphicx} 
\usepackage{dcolumn}
\usepackage{bm}
\usepackage{amssymb}  
\usepackage{cite}
\usepackage{amsthm} 
\usepackage[letterpaper,top=2cm,bottom=2cm,left=3cm,right=3cm,marginparwidth=1.75cm]{geometry}
\usepackage{amsmath} 
\usepackage[colorlinks=true, allcolors=blue]{hyperref}
\usepackage{authblk}
\usepackage{tabularx}
\usepackage{xcolor}
\usepackage{threeparttable}
\usepackage{longtable}
\usepackage{array}
\usepackage{caption}
\usepackage{ragged2e} 
\usepackage{multirow} 

\newcolumntype{L}[1]{>{\raggedright\let\newline\\\arraybackslash\hspace{0pt}}m{#1}}
\newcolumntype{C}[1]{>{\centering\let\newline\\\arraybackslash\hspace{0pt}}m{#1}}

\begin{document}
	\title{Domain matters: Towards domain-informed evaluation for link prediction}
	\author{Yilin Bi, Junhao Bian, Shuyan Wan, Shuaijia Wang, Tao Zhou$^{*}$}
	\affil{CompleX Lab, School of Computer Science and Engineering, University of Electronic Science and Technology of China, Chengdu 610054, China. \\ $^{*}$Corresponding author: zhutou@ustc.edu}
	\maketitle

\begin{abstract}
Link prediction, a foundational task in complex network analysis, has extensive applications in critical scenarios such as social recommendation, drug target discovery, and knowledge graph completion. However, existing evaluations of algorithmic often rely on experiments conducted on a limited number of networks, assuming consistent performance rankings across domains. Despite the significant disparities in generative mechanisms and semantic contexts, previous studies often improperly highlight ``universally optimal" algorithms based solely on naive average over networks across domains. This paper systematically evaluates 12 mainstream link prediction algorithms across 740 real-world networks spanning seven domains. We present substantial empirical evidence elucidating the performance of algorithms in specific domains. This findings reveal a notably low degree of consistency (correlation $\approx 0.0991$) in inter-domain algorithm rankings, a phenomenon that stands in stark contrast to the high degree of consistency observed within individual domains. Principal Component Analysis (PCA) shows that response vectors formed by the rankings of the 12 algorithms cluster distinctly by domain in low-dimensional space, thus confirming domain attributes as a pivotal factor affecting algorithm performance. We propose a metric called Winner Score that could identify the superior algorithm in each domain, which suggest the following domain-specific winners: Non-Negative Matrix Factorization (NMF) for social networks, Neighborhood Overlap-aware Graph Neural Networks (NeoGNN) for economics, Graph Convolutional Networks (GCN) for chemistry, and L3-based Resource Allocation (RA3) for biology. However, these domain-specific top-performing algorithms tend to exhibit suboptimal performance in other domains. This finding underscores the importance of aligning an algorithm's mechanism (\textit{e.g.}, low-rank modeling, high-order path propagation) with the network structure. In addition, the proposed Ranking Stability Coefficient (RSC) --which quantifies the number of networks required for stable evaluation--reveals a significant discrepancy in the stability requirements across different domains. Highly homogeneous domains (\textit{e.g.}, transportation, information) achieve stability with approximately 10 networks, whereas highly heterogeneous domains (\textit{e.g.}, biology, economics) require more than 75 networks. This study provides empirical evidence for algorithm selection, benchmark construction, and reproducible evaluation, thereby advancing link prediction from general comparison to scenario adaptation.
\end{abstract}
\vspace {2.5mm}
 \par

\section{Introduction}
From social networks \cite{freeman2004, vega2007, malley2008} to biological pathways \cite{barabasi2004, gosak2018}, from economic collaborations \cite{foster2005} to traffic flows \cite{cascetta2009}, numerous real-world systems, despite their diverse contexts, share a fundamental structural representation: Modeling their interaction structures through complex networks composed of nodes and their links \cite{holling2001, mitchell2009, chen2015, barabasi2016, newman2018, xu2025}. Despite commonalities in the fundamental form of graph structures, significant disparities emerge in the underlying generation mechanisms, evolutionary dynamics, and functional constraints of these inter-domain networks upon more detailed examination. Social networks, for instance, manifest pronounced homogeneity and community structures \cite{granovetter1973, newman2003}. Protein interaction networks, by contrast, feature modular organization and path propagation properties \cite{jeong2000, jeong2001}. Economic cooperation networks are influenced by institutional rules and sparse interactions \cite{acemoglu2015}, while technological infrastructure emphasizes robustness and hierarchical design \cite{doyle2005}.
	
Link prediction, a fundamental component of network analysis, aims to identify unobserved but potentially existing links or forecast emerging links \cite{lv2011, wang2015, martinez2016, kumar2020, zhou2021, wu2022, arrar2024}. This task not only serves as a vital tool for understanding network formation mechanisms but also plays practical roles in multiple critical scenarios, including friend recommendations on social platforms \cite{leskovec2010}, gene function inference in biological networks \cite{musawi2025}, and completing missing relationships in knowledge graphs \cite{kazemi2018}. The effectiveness of algorithms in real-world scenarios must ultimately be evaluated by their performance on actual networks. A review of link prediction reveals that the field has evolved from constrained to broader contexts. In the early stages of this field, researchers faced significant challenges in data acquisition and computational resources access. To address these issues, they frequently employed a set of curated classic ``benchmark datasets''. These datasets, including Zachary's Karate Club \cite{zachary1977}, Power \cite{watts1998}, Jazz \cite{gleiser2003}, and CollegeMsg \cite{panzarasa2009}, were used to validate researchers' methods. These datasets played a crucial role in the field's infancy by providing reproducible testing environments for algorithm design. However, relying on these networks for performance evaluation introduces inherent limitations, including limited diversity (primarily citations and social networks), small scale, and ``friendly" topological features (high density, strong clustering). Despite the availability of vast real-world networks shared through public platforms (\textit{e.g.}, Network Repository \cite{rossi2015}, KONECT \cite{kunegis2013}), many studies persist in the early-formed habit of conducting experiments solely on these underrepresented samples. Claims of broad applicability are made for novel methodologies, yet the heterogeneity and intricacy of actual networks are disregarded. This evaluative framework, established during an era of data scarcity, falls to reflect the genuine efficacy of algorithms across disparate domains. Moreover, it carries the potential to misinterpret local advantages as universal superiority.
	
More critically, even when studies incorporate more real-world networks, their evaluation methods remain significantly flawed. Some research tests algorithms on multiple networks within a single domain (\textit{e.g.}, social or biological) and then asserts their applicability to all network types. Others span multiple domains but ignore contextual differences in analysis, simply averaging performance metrics across domains to select an ``universally optimal'' algorithm. These practices obscure a long-neglected fundamental question: \textbf{When evaluating link prediction algorithms across different domains, are the domain-specific rankings of algorithms consistent and stable?} If an algorithm's relative strengths vary depending on the network's domain, conclusions based on local data or simple averaging cannot be directly generalized to other domains.
	
Notably, despite the lack of systematic research addressing this issue, a few studies suggested that the algorithms’ performance is domain-dependent. For instance, Zhou \textit{et al.} \cite{zhou2021b} compared the performance of local similarity metrics based on second-order and third-order paths across 128 real networks in 16 domains. The researchers found that third-order metrics outperformed second-order ones in software networks and food chain networks, while second-order metrics were notably effective in collaboration networks, human contact networks, and animal contact networks. Within the context of protein-related networks, third-order metrics demonstrated excellence in protein interaction networks but exhibited suboptimal performance in protein structural networks, where nodes represented secondary structural units. Muscoloni \textit{et al.} \cite{muscoloni2023} conducted a comparative analysis of algorithms including CHA \cite{carlo2022}, SPM \cite{lv2015}, SBM \cite{roger2009}, and Stacking \cite{ghasemian2020} across 550 networks in six domains. The results indicate: CHA achieved the highest win rate across diverse networks (biological, informational, technological, economic, and social domains) but performed relatively poorly in transportation networks. In contrast, SPM excelled in transportation networks with the highest win rate, while consistently ranking second in the remaining five categories. Additionally, SBM has exhibited a notable performance link over Stacking in biological, informational, and economic networks. However, its performance is found to be deficient in technological, social, and transportation networks. In a related study, He \textit{et al.} \cite{he2024} made a notable discovery: In social networks, a predictor constructed using a low-order link trajectory equivalent to the Common Neighbor heuristic attains a $91\%$ win rate in pairwise comparisons against other predictors based on different trajectory structures. Nevertheless, in biological, technological, and economic networks, the win rate of the same predictor was consistently below $10\%$, indicating that its effectiveness is highly dependent on the domain-specific characteristics of the network. The scattered yet consistent occurrence of these phenomena suggests that the performance of link prediction algorithms is not inherent to the algorithms themselves but rather a result of the interaction between the algorithm and the network, leading to varying performance across different domains. This finding calls into question the widely held belief that ``a universally optimal algorithm exists'' and underscores the limitations of prevailing evaluation frameworks when applied to real and diverse networks.
	
 In light of these considerations, a critical yet frequently overlooked issue directly impacting experimental reliability demands immediate attention: \textbf{When conducting link prediction for networks in a specific domain, how many real-world networks from that domain must be included to ensure reliable algorithm performance evaluation—that is, to guarantee stable algorithm rankings?} Insufficient samples may result in conclusions disproportionately influenced by individual anomalous networks, thereby introducing evaluation bias. Conversely, an indiscriminate expansion of the number of networks may necessitate unnecessary data processing and computational overhead. Currently, this fundamental issue concerning the rationality of experimental design \cite{lakens2022} has not been systematically explored.

To address this gap, this paper systematically examines the evaluation stability of link prediction algorithms across domains. A total of 740 real-world networks spanning seven domains—biological, chemical, economic, informational, social, technological, and transportation—were collected. These networks were then comprehensively evaluated using 12 mainstream algorithms. The performance distributions, ranking consistency, and win rates of these algorithms were analyzed to reveal how algorithmic advantages shift across different domains. In the context of this study, the number of networks required to achieve stable performance rankings across domains is quantified. This provides quantifiable and reproducible operational guidelines for selecting network sample sizes within specific domains. This study challenges the simplistic assumption of a ``universally optimal'' algorithm. It also provides methodological support for constructing refined evaluations grounded in real-world data diversity. The remainder of this paper is organized as follows: Section 2 provides a comprehensive definition and fundamental information regarding link prediction. Section 3 details the primary study findings. Section 4 offers a summary of the paper and delineates prospective avenues for future research.

\section{Preliminaries}
Consider a simple undirected, unweighted network $G(V, E)$ containing no self-loops or multiple links. The node set is denoted by $V = \{v_1, v_2, ..., v_n\}$, and the link set is denoted by $E = \{e_1, e_2, ..., e_m\}$. Let $U$ denote the universal set of all possible latent links, defined as: $$U=\{\{v_i, v_j\} \mid v_i, v_j \in V, i<j\}.$$
The size of the potential link space is defined as $|U|=\frac{|V|(|V|- 1)}{2}$, representing the number of potential links in the network. The objective of link prediction is to identify genuinely existing but missing links from the unobserved link set $U-E$. However, given that links in $U-E$ may correspond to either genuinely missing links or nonexistent links, their true labels remain unknown. To evaluate the accuracy of the algorithm, the standard protocol is to randomly partition the observed link set $E$ into a training set $E^T$, a validation set $E^V$, and a probe set $E^P$ \cite{jiao2025}, satisfying the following relations:
$$ E^T \cup E^V \cup E^P = E, \quad E^T \cap E^P = E^T \cap E^V = E^V \cap E^P = \emptyset. $$
During training, $E^T$ is exclusively used for building the prediction model, while $E^P$ (referred to as the probe set) is treated as positive samples, that is, those that are genuinely missing links. For algorithms requiring negative samples, a straightforward approach is employed, involving randomly selecting $|E^P|$ links from $U - E$ as negative samples (\textit{i.e.}, non-existent links). Despite the potential of more sophisticated negative sampling strategies (\textit{e.g.}, topology-based) to enhance prediction accuracy, the impact of negative sampling strategies remains a subject \cite{deng2025} that is not addressed in this paper.
For parameterized algorithms, the model is trained on the training set $E^T$. The performance of different hyperparameter configurations is evaluated on the validation set $E^V$. After selecting the optimal configuration, we merge $E^T$ and $E^V$ to retrain the final model. For non-parameterized algorithms, $E^T$ and $E^V$ are directly merged to construct the predictor. Each link prediction algorithm generates a likelihood score $s_{ij}$ for each node pair $(v_i, v_j)$ where $i,j \in V$ and $(v_i, v_j) \in E^T\cup E^V$. This score is then used to rank potential links. Ideally, positive samples should receive higher scores, and negative samples lower scores. This study adopts the split ratio $E^T : E^V : E^P = 0.64 : 0.16 : 0.20$. Twelve representative link prediction algorithms were selected for evaluation, covering various methodologies including local heuristics, global embedding methods, and matrix factorization approaches. Detailed descriptions of each algorithm can be found in \autoref{tab:algorithms}.

\begin{table}[htbp]
	\centering
	\caption{\textbf{Summary of link prediction algorithms.} 
		The hyperparameter settings of each algorithm are as follows: For DW, the random walk length is set to 40 and 80, the number of random walks per node is 10 and 80, the embedding vector dimensions include three options: 32, 64, and 128, and the window size in Word2Vec is 5 and 10; the embedding vector dimension of NMF is set to 32, 64, and 128; the hidden layer vector dimension of GCN is 32, 64, and 128, while the hidden layer vector dimensions of other related algorithms are respectively GAT=16, GraphSAGE=256, GAE=128, VGAE=128, and NeoGNN=128. It should be noted that for hyperparameters with multiple optional values, the optimal parameters are selected through traversal: First, the model is trained on the training set, and the performance of the model under different parameter combinations is compared on the validation set. The parameters corresponding to the highest AUC score on the validation set are selected as the final model parameters, and then the validation set is merged into the training set for retraining. MFI used a fixed parameter $\alpha = 0.1$. }
	\label{tab:algorithms}
	\renewcommand{\arraystretch}{1.4}
	\begin{tabularx}{\textwidth}{@{}>{\RaggedRight}p{12cm}  >{\RaggedRight}X l@{}}
		\toprule
		\textbf{Algorithm} & \textbf{Reference} \\
		\midrule
		DeepWalk (DW) 
		& \cite{perozzi2014} \\
		
		Graph Auto-Encoder (GAE)
		& \cite{kipf2016} \\
		
		Graph Attention Networks (GAT)
		& \cite{velickovic2018} \\
		
		Graph Convolutional Network (GCN) 
		& \cite{chen2020a} \\
		
		Sample and aggreGatE (GraphSAGE)
		& \cite{hamilton2017} \\
		
		Matrix Forest Index (MFI) 
		& \cite{chebotarev1997} \\
		
		Neighborhood Overlap-aware GNNs (NeoGNN)
		& \cite{yun2021} \\
		
		Non-negative Matrix Factorization (NMF) 
		& \cite{chen2020b} \\
		
		Propagating Uncertainty through Latent Links (PULL)
		& \cite{kim2025} \\
		
		Resource Allocation (RA) 
		& \cite{ou2007, zhou2009} \\
		
		L3-based Resource Allocation (RA3) 
		& \cite{kovacs2019, pech2019} \\
		
		Variational Graph Auto-Encoder (VGAE)
		& \cite{kipf2016} \\
		
		\bottomrule
	\end{tabularx}
\end{table}

 \section{Results}
\begin{figure}[H]
	\centering
	\includegraphics[width=1.0\linewidth]{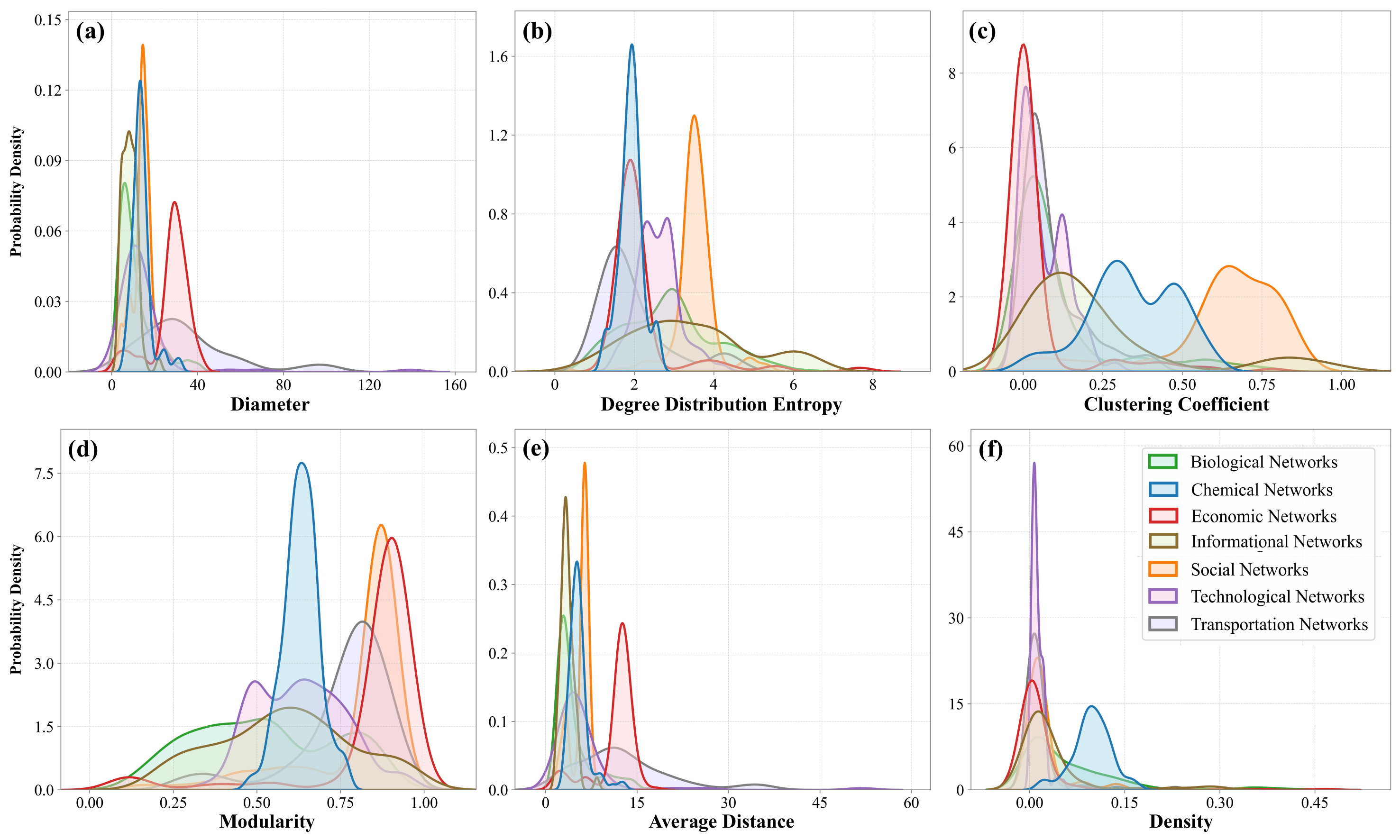}
	\caption{{\bf Distributions of topological features of networks in different domains.} 
		(a)--(f) show kernel density estimates of six topological features, with detailed definitions and mathematical formulations as follows: 
		(a) \textbf{Diameter}: The longest distance between any pair of nodes in the largest connected component (LCC) of the graph, defined as: $\Delta(G) = \max_{u, v \in V_{\mathrm{lcc}},\, u \neq v} d(u, v)$, where $V_{\mathrm{lcc}}$ denotes the node set of the LCC and $d(u, v)$ represents the distance (\textit{i.e.}, length of the shortest path) between nodes $u$ and $v$; 
		(b) \textbf{Degree Distribution Entropy}: A measure to quantify the heterogeneity of the node degree distribution, calculated as $-\sum_{k: P(k) > 0} P(k) \log_2 P(k)$ with $P(k) = n_k / n$ is the probability that a randomly selected node has degree $k$, $n_k$ is the number of nodes with degree $k$; 
		(c) \textbf{Clustering Coefficient}: The ratio of three times the number of triangles to the number of connected triples; 
		(d) \textbf{Modularity}: If evaluates  the quality of community partitioning $ \mathcal{P}=\{C_1, C_2, \dots, C_k\}$, computed as $Q(G, \mathcal{P}) = \sum_{i=1}^{k} \left[\frac{l_i}{m} - \left( \frac{d_i}{2m} \right)^2 \right]$, where $l_i$ is the number of links within community $C_i$, $d_i$ is the total degree of nodes in $C_i$; 
		(e) \textbf{Average Distance}: The average distance between all pairs of nodes in the largest connected component of $G$: $ \frac{1}{|V_{\mathrm{lcc}}|(|V_{\mathrm{lcc}}| - 1)} \sum_{u \neq v \in V_{\mathrm{lcc}}} d(u, v)$; 
		(f) \textbf{Density}: The proportion of actual links to all possible links in the graph, calculated as $2m / [n(n - 1)]$. Each curve in the figures corresponds to networks from one domain.}
	\label{fig_data}
\end{figure}

 This study collected 740 real-world network datasets from seven application domains: 185 biological networks, 100 chemical networks, 127 economic networks, 55 informational networks, 138 social networks, 70 technological networks, and 65 transportation networks. The datasets were primarily sourced from two public repositories: Network Repository \cite{rossi2015} and KONECT \cite{kunegis2013}. \autoref{fig_data} illustrates the distribution of these networks across six key topological properties, reflecting their structural heterogeneity. For instance, social networks are known to exhibit high clustering coefficients, whereas technological networks typically have low link density. These inter-domain structural differences provide an empirical basis for subsequent analysis of how link prediction algorithm performance varies across data sources.

Building on recent studies on link prediction evaluation frameworks \cite{zhou2023, jiao2024, bi2024, wan2025}, this paper adopts Area Under the ROC Curve (AUC) \cite{hanley1982} as the primary performance metric. AUC quantifies a model's ability to distinguish positive samples (missing links) from negative samples (nonexistent links) across all possible classification thresholds. Its theoretical range is $[0,1]$, where higher values indicate better prediction performance. From a probabilistic perspective, AUC is equivalent to the probability that a randomly selected positive sample will have a higher score than a randomly selected negative sample. In this study, we employed a random comparison method, selecting $N$ pairs of positive and negative samples for analysis. If in $N_1$ of these pairs the missing link has a higher score and in $N_2$ pairs the missing link and nonexistent link have the same score, then AUC is approximately equal to:
\begin{equation}
		 \mathrm{AUC} \approx \frac{N_1 + 0.5 N_2}{N}.
\end{equation}
When the algorithm produces completely random predictions, AUC approaches 0.5. Consequently, the extent to which AUC exceeds 0.5 indicates the algorithm's performance advantage over random guessing. All results reported in this paper are the average AUC values obtained from ten independent random splitting experiments. The results for other evaluation metrics are detailed in the Appendix and are consistent with those of AUC.

\subsection{Correlations of algorithm rankings}
\begin{figure}[t]
	\centering
	\centerline{\includegraphics[width=1.0\linewidth]{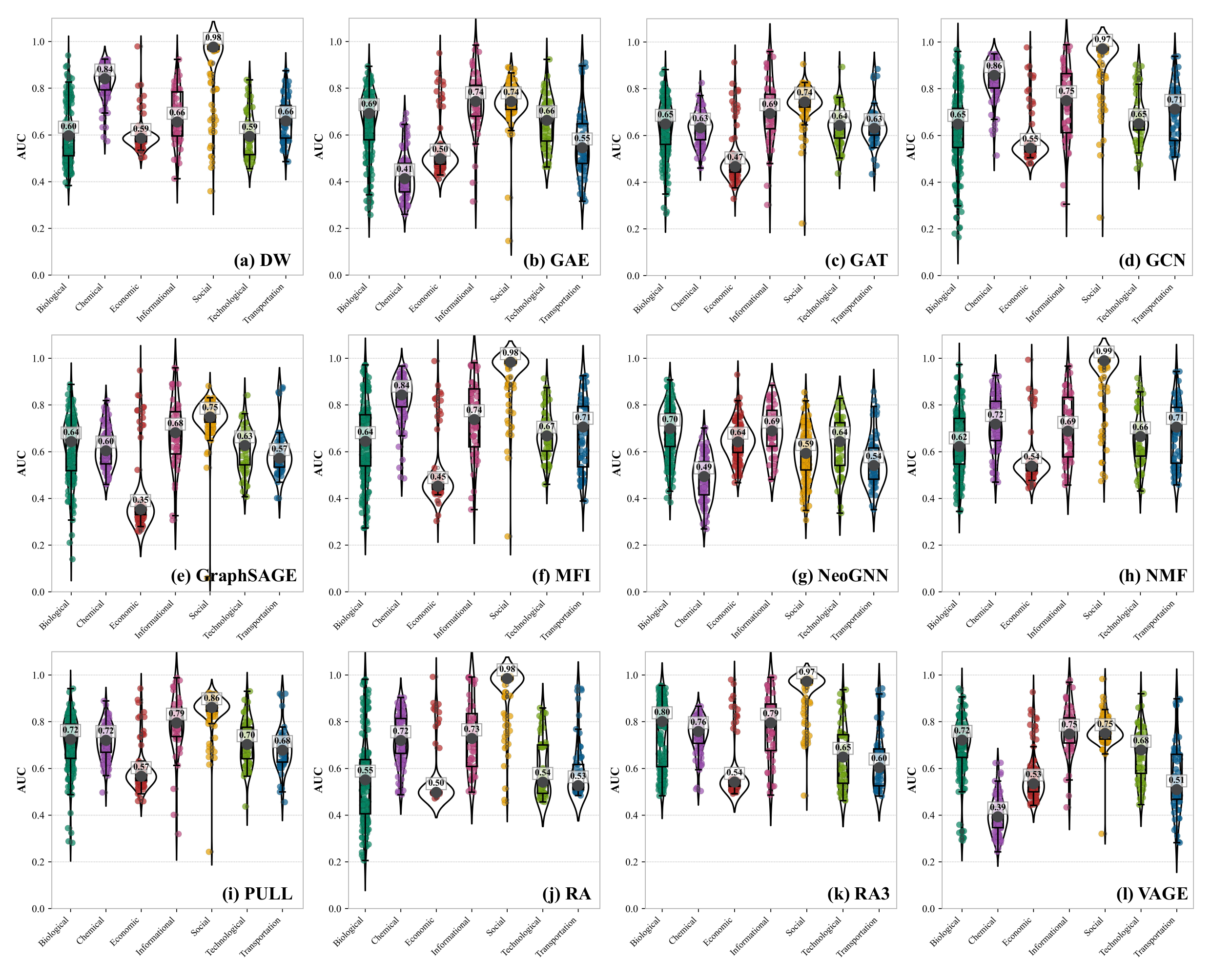}}
	\caption{{\bf Performance comparison of the 12 link prediction algorithms across the seven domains.} Box plots illustrate the distribution of AUC scores achieved by each of the 12 link prediction algorithms on networks from the seven domains. The x-axis represents domain, while the y-axis denotes the AUC scores. The central mark within each box indicates the median of the AUC score distribution.}
	\label{fig_box}
\end{figure}

To examine the stability of link prediction algorithms across different domains, we evaluated the performance of the 12 classical link prediction algorithms mentioned above on 740 real networks across seven domains. As shown in \autoref{fig_box}, the AUC values of each algorithm vary substantially across domains, indicating a correlation between the domain of a network and the prediction performance of algorithms on it. Specifically, in social networks, most algorithms exhibit generally higher AUC values with narrower distribution ranges and lower dispersion, suggesting that the topological structure of such networks is relatively consistent in its response to different prediction mechanisms. Conversely, algorithmic performance in economic networks is generally suboptimal. Despite the relative stability in algorithm rankings (\textit{i.e.}, high intra-domain consistency), the absolute predictive performance of these algorithms remains limited. This phenomenon may stem from factors such as the complexity of link formation mechanisms or data sparsity within these networks. Biological networks exhibit highly heterogeneous algorithmic performance distributions: AUC values vary significantly across different networks within the same domain, with standard deviations exceeding those of other domains. This finding indicates substantial differences in the mechanisms and patterns underlying the formation of distinct network structures within biological systems. In contrast, chemical networks demonstrate relatively concentrated performance distributions, with average AUC values at moderate levels. Transportation networks exhibit moderate average performance but greater dispersion, approaching the dispersion levels of biological networks.

These discrepancies in inter-domain performance underscore the notion that \textbf{ the superior performance of an algorithm in one specific domain does not guarantee its direct applicability to networks in other domains}. In other words, domain attributes act as implicit structural moderators, systematically influencing the predictive performance of algorithms. Consequently, the claim of a universally optimal algorithm for link prediction is unsubstantiated \cite{ghasemian2019}. This finding further underscores the necessity of quantifying intra-domain algorithmic ranking consistency-precisely the motivation for introducing Kendall's $\tau$ ranking consistency analysis in subsequent sections.
	
 A critical question thus arises: Do the performance rankings of algorithms remain consistent across networks within the same domain? Answering this question helps quantify the similarity of algorithm performance rankings both within and across domains. To this end, we adopt Kendall's $\tau$ correlation coefficient \cite{kendall1938} as our metric. The $p$-th domain contains $u_p$ networks, and its algorithm performance ranking matrix is denoted as $D^p \in \mathbb{R}^{u_p \times m}$, where $m$ is the number of algorithms (\textit{i.e.}, $m=12$). The element $d_{il}^p$ represents the performance ranking of the $l$-th algorithm among all $m$ algorithms on the $i$-th network in the $p$-th domain. The algorithms are ranked in descending order of AUC, with the top-performing one ranked first. Kendall's $\tau$ is a nonparametric statistic used to measure the degree of agreement between two total orderings of the same set of objects (\textit{i.e.}, algorithm rankings). For two ranking vectors $\bm{a}$ and $\bm{b}$ of length $m$ (corresponding to algorithm rankings on two networks), Kendall's $\tau$ takes values in $[-1,1]$: $\tau = 1$ indicates perfect agreement, $\tau = -1$ indicates perfect disagreement, and $\tau = 0$ indicates no linear relationship. Its classical form is defined as: $$\tau(\bm{a}, \bm{b}) = \frac{N_c - N_d}{\binom{m}{2}},$$ where $N_c$ and $N_d$ denote the number of concordant pairs and discordant pairs, respectively. A pair of indices $(j, k)$ with $j<k$ is called concordant if $(a_j - a_k)(b_j - b_k) > 0$; otherwise, it is discordant.
	
Therefore, for any two domains, $p$ and $q$, the \textbf{inter-domain algorithm ranking consistency} is defined as the average correlation coefficient of algorithm ranking sequences across all inter-domain network pairs:
\begin{equation}
	R^{pq} = \frac{1}{u_p u_q} \sum_{i=1}^{u_p} \sum_{j=1}^{u_q} \tau\left( \bm{d}_i^p, \bm{d}_j^q \right),
\end{equation}
 where $\bm{d}_i^p = [d_{il}^p, d_{i2}^p, \dots, d_{im}^p]$ denotes the algorithm ranking vector for the $i$-th network. Correspondingly, the \textbf{intra-domain algorithm ranking consistency} for the $p$-th domain is defined as the average Kendall's $\tau$ correlation coefficient across all pairwise networks within that domain:
\begin{equation}
	R^{pp} = \frac{1}{u_p(u_p - 1)} \sum_{i=1}^{u_p} \sum_{\substack{j=1 \\ j \neq i}}^{u_p} \tau\left( \bm{d}_i^p, \bm{d}_j^p \right).
\end{equation}
These two metrics respectively reflect the stability of algorithmic ranking behavior within specific domains and the inter-domain transferability potential of algorithmic advantages.
	
\begin{figure}[htbp]
	\centering
	\centerline{\includegraphics[width=0.90\linewidth]{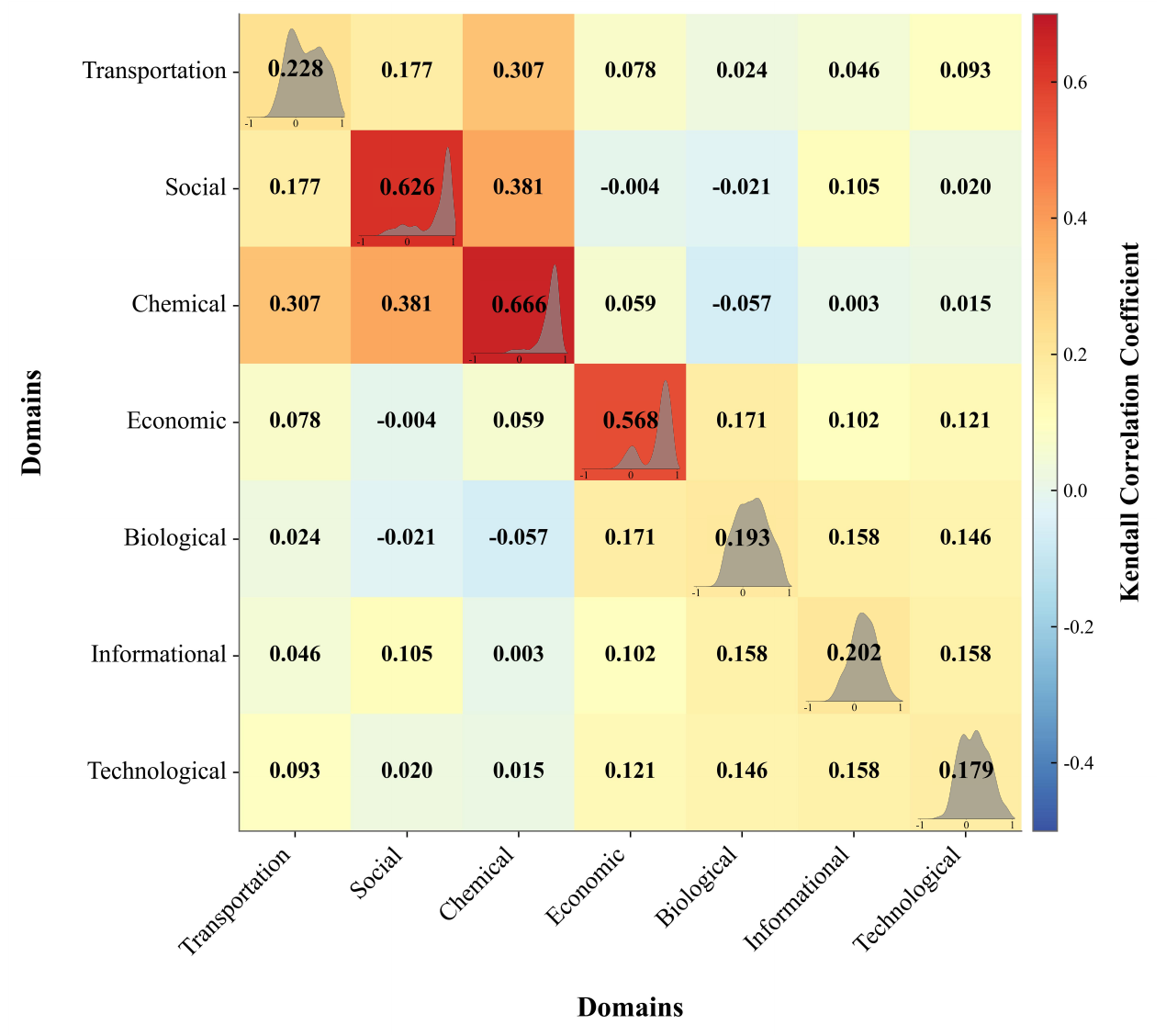}}
	\caption{{\bf Correlation of algorithm rankings across different domains.} Heatmap depicting the Kendall's $\tau$ correlation coefficients of algorithm rankings between and within domains.}
	\label{fig_kcc}
\end{figure}

 As demonstrated in \autoref{fig_kcc}, within chemical, social, and economic networks, there is a high degree of intra-domain ranking consistency ($R^{pp} > 0.5$) with concentrated distributions. This finding suggests stable algorithmic performance in prioritizing superior nodes across these network types. In contrast, the $R^{pp}$ values for biological and transportation networks are significantly lower, with dispersed distributions. This indicates substantial variation in algorithmic rankings across different networks within the same domain, reflecting the complexity of their internal structural formation mechanisms.
	
Notably, the inter-domain algorithm ranking consistency is generally weak: The average $R^{pq}$ across all inter-domain pairs is $0.0991$, which is significantly lower than the average intra-domain $R^{pp}$ of $0.4117$. The diagonal elements of the heatmap, which correspond to $R^{pp}$ (intra-domain ranking consistency), represent the maximum values in each row and column. This indicates that the similarity of algorithm rankings within a specific domain is higher than that with other domains. However, even along the diagonal, the correlation coefficients remain predominantly moderate (typically below $0.6$), indicating that while consistency exists within domains, it is not absolute. Furthermore, the majority of off-diagonal elements approach zero, indicating that the relative advantages of current mainstream link prediction algorithms lack inter-domain transferability. Specifically, certain domain pairs (\textit{e.g.}, social--economic, social--biological, biological--chemical) exhibit weak negative correlations ($R^{pq}<0$), suggesting significant differences in their underlying network generation mechanisms or edge formation mechanisms. This further confirms that algorithms performing well in one domain may exhibit suboptimal performance in another.

\subsection{Domain-specific competitive landscapes of algorithms}
Differences in structural generation mechanisms and functional characteristics across domains may influence the relative performance of link prediction algorithms within them. To identify dominant algorithms in each domain, we introduce the \textbf{Winner Score} metric, defined as the weighted sum of high rankings achieved by an algorithm across all networks within a specific domain. The $p$-th domain contains $u_p$ networks. The weighted score of the $l$-th algorithm in this domain is:
\begin{equation}
		 W_l^p = \frac{1}{u_p} \sum_{i=1}^{u_p} \frac{1}{d_{il}^p},
\end{equation}
where $d_{il}^p$ denotes the performance ranking of algorithm $l$ on network $i$ within domain $p$. Rankings are determined by sorting AUC values in descending order across all 12 algorithms on that network (highest AUC ranks 1st). This metric assigns higher weights to top rankings: An algorithm contributes 1 to the sum when ranked first on a network, with contributions decreasing as rankings drop. Thus, a larger $W_l^p$ indicates stronger overall competitiveness of algorithm $l$ in domain $p$. To facilitate inter-domain comparisons, we normalize $W_l^p$ and define the normalized winner score of algorithm $l$ in domain $p$ as:
\begin{equation}
		 \hat{W}_l^p = \frac{W_l^p}{\sum_{r=1}^{12}W_{r}^{p}}.
\end{equation}

\begin{figure}[htbp]
	\centering
	\centerline{\includegraphics[width=0.95\linewidth]{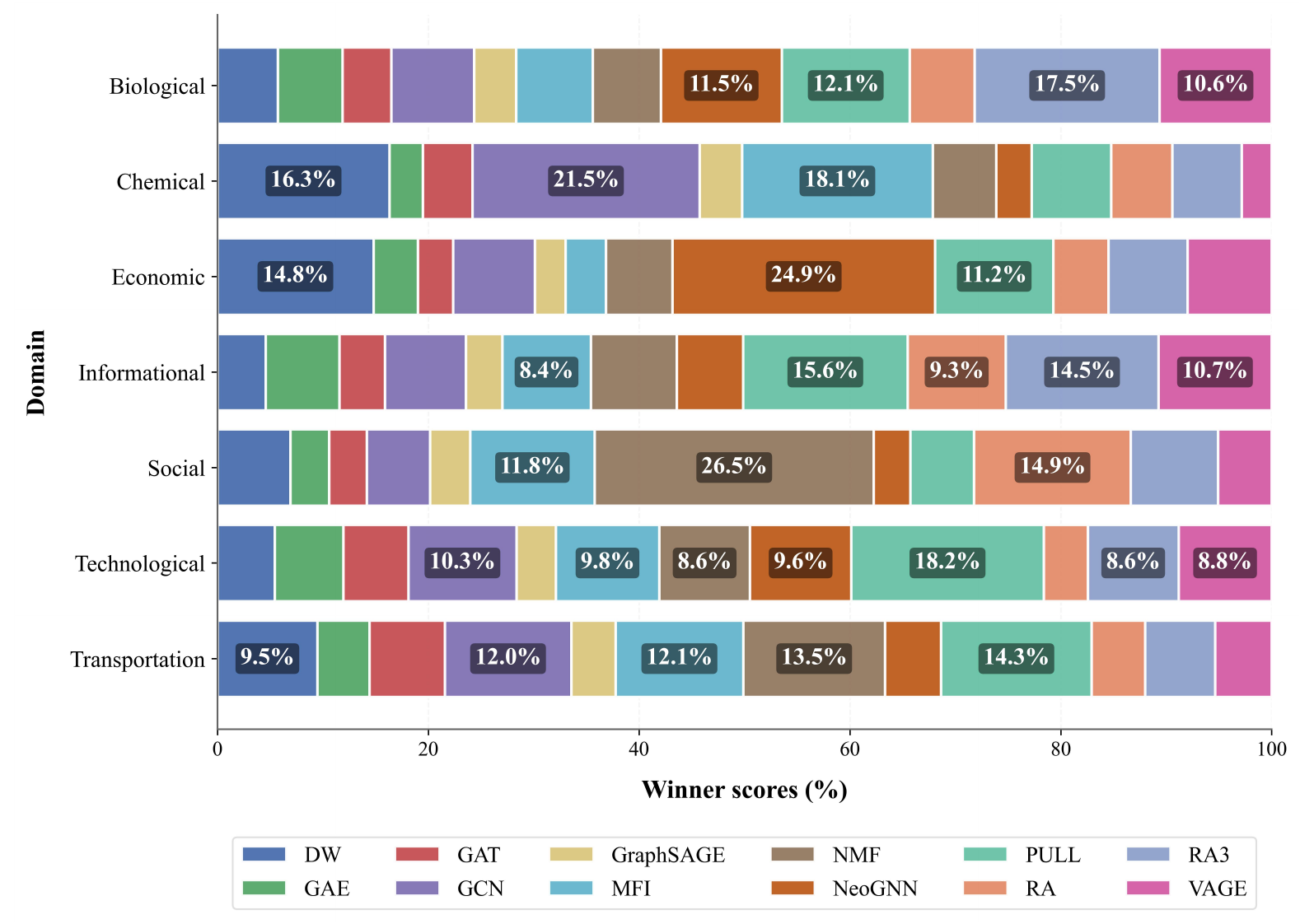}}
	\caption{{\bf Winner scores of the 12 link prediction algorithms across different domains.} Stacked bar chart illustrates the winner score of each of the 12 link prediction algorithms across the seven network domains. The y-axis represents the network domains, while the x-axis denotes the cumulative percentage of the winner score within each domain. The length of each colored segment indicates the winner score of the corresponding algorithm. Black boxes highlight instances where the winner score exceeds the average chance level ($100\%/12 \approx 8.33\%$), signifying that the algorithm’s performance is superior to the average level.}
	\label{fig_winnerrate}
\end{figure}

As shown in \autoref{fig_winnerrate}, \textbf{the distribution of algorithmic advantages exhibits pronounced domain dependence, and no universal algorithm achieves the highest Winner Score across all domains}. Specifically: In social networks, the NMF algorithm achieves a leading score of 26.5\%, while in economic networks, NeoGNN emerges as the dominant approach with 24.9\%. Similarly, within chemical networks, GCN secures the top position with 21.5\%, and in biological networks, the RA3 algorithm outperforms other algorithms with 17.5\%. These results imply that the inherent structural biases of different algorithms may exhibit varying degrees of congruence with the topological characteristics of specific domains. For instance, NMF captures global structural homogeneity through low-rank matrix decomposition, thus demonstrating a notable advantage in social networks with strong community structures. NeoGNN exhibits superior performance in sparsely connected economic networks due to its graph neural network co-occurrence mechanism that integrates neighbor information. The GCN model leverages message passing to model local topological adjacency, thus demonstrating competitiveness in chemical networks with well-defined atomic-level topological connectivity. On the other hand, the RA3 algorithm leverages odd-degree adjacency relationships, proving more effective in bipartite biological networks \cite{petter2003} with strong bipartiteness.
	
Further analysis of inter-domain algorithmic competition reveals distinct patterns: In social networks, NMF and in economic networks, NeoGNN achieve Winner Scores significantly higher than the next-best algorithm ($\geq 10$ percentage points), indicating strong domain-specific preferences. In contrast, in chemical networks, Winner Scores for GCN, MFI, and DW differ by less than 2 percentage points, reflecting comparable performance across different modeling paradigms. While PULL is optimal in transportation networks, the Winner Scores of other algorithms exhibit a wide distribution with no distinct performance cutoff. It is important to note that \textbf{an algorithm that performs optimally in one specific domain may not necessarily exhibit a significant performance advantage in other domains}. For instance, NMF exhibits strong performance in social networks but exhibits suboptimal performance with an average Winner Score of 8.33\% across informational, economic, chemical, and biological networks. Similarly, RA3 exhibits notable performance in biological networks but exhibits mediocre performance in social, transportation, economic, and chemical networks. This phenomenon suggests that an algorithm's effectiveness depends on the congruence between its intrinsic properties (\textit{e.g.}, locality, low-rank property, high-order transitivity) and the topological characteristics of the target network, rather than its model complexity alone. Therefore, the selection of link prediction algorithms for practical applications should not assume a ``universally optimal'' algorithm but rather should adopt a mechanism-adaptive approach based on domain-specific topological characteristics. This finding further undermines the validity of extrapolating algorithmic performance advantages from one domain to another without considering the specific application context.

 \subsection{Domain-specific clusters in networks' low-dimension representation}
\begin{figure}[htbp]
	\centering
	\centerline{\includegraphics[width=1.0\linewidth]{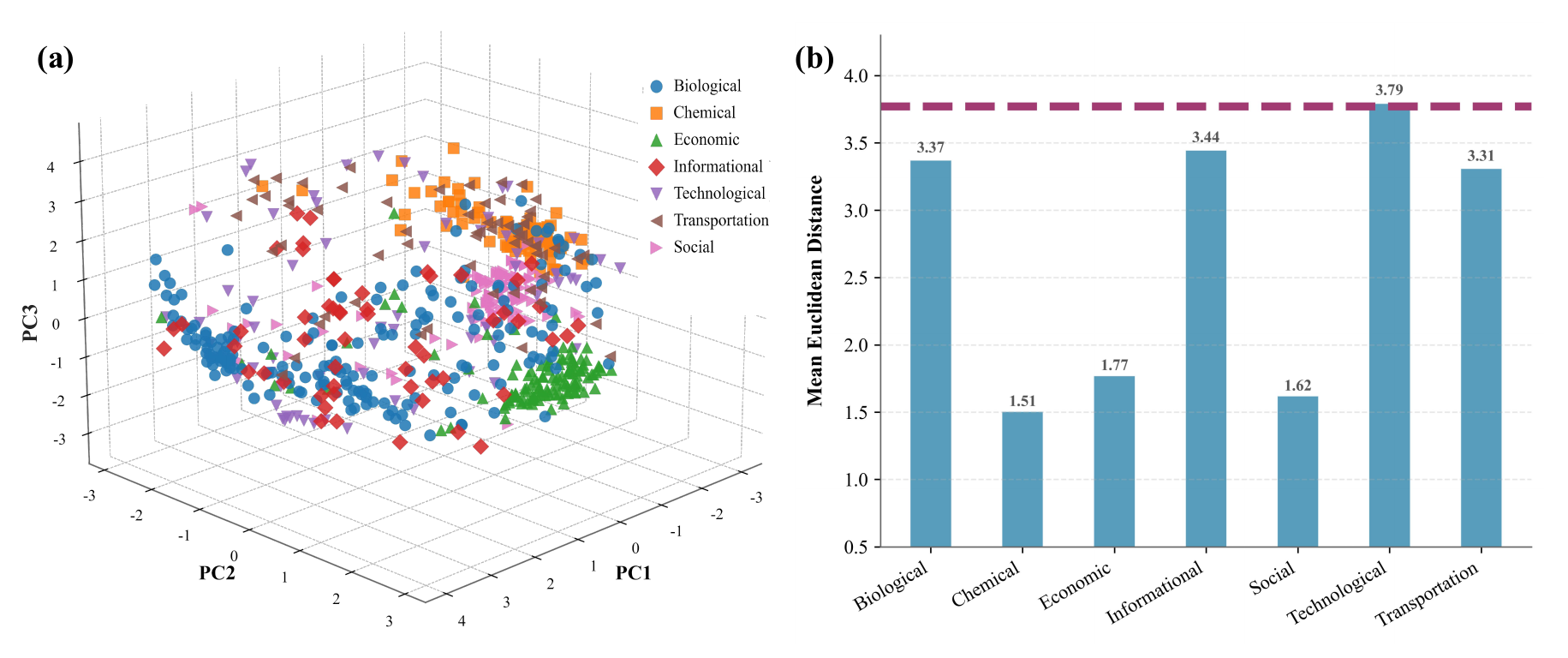}}
	\caption{{\bf  Principal component analysis (PCA) of algorithm rankings across domains.} (a) 3D scatter plot where each point represents a network, with coordinates defined by the first three principal components (PC1, PC2, and PC3). Different colors or markers are used to distinguish networks from distinct domains. (b) Comparison between actual intra-domain distances and those obtained via permutation tests. The blue y-axis denotes the value of $\bar{D}_{p}$, while the pink dashed line denotes $\bar{D}_{\text{rand}}$, whose value is $3.77$.}
	\label{fig_pca}
\end{figure}

 The preceding analysis demonstrates significant variations in link prediction algorithm performance across domains, which prompts the following question: Do these discrepancies stem from identifiable underlying patterns? We conduct Principal Components Analysis (PCA) \cite{dunteman1989} on the performance ranking vectors of all networks. Each network is represented as a 12-dimensional vector, where each dimension corresponds to the performance ranking of one of the 12 link prediction algorithms. The $j$-th dimension of this vector represents the performance ranking $d_{ij}^p$ of the $j$-th algorithm on that network. In other words, the PCA input space is a high-dimensional space constructed from algorithm performance rankings, where each dimension corresponds to the relative performance ranking of one algorithm across all networks. To illustrate, if in a social network, NMF is ranked first, GCN third, and RA3 fifth (based on AUC values), its feature vector values on the NMF, GCN, and RA3 dimensions would be 1, 3, and 5, respectively. This representation abstracts networks into ``algorithmic performance fingerprints'', where clustering reflects consistency in domain-specific algorithm preference patterns. The PCA results show that the first three principal components cumulatively explain 70.8\% of the total variance (PC1: 39.16\%, PC2: 16.71\%, PC3: 14.93\%), suggesting that the performance patterns of the 12 link prediction algorithms can be effectively compressed into a low-dimensional space. After projecting each network onto the three-dimensional embedding space (see \autoref{fig_pca}(a) for details on the PCA procedure), networks from the same domain tend to cluster together. This tendency is characterized by smaller intra-domain network distances and larger inter-domain network distances. This finding underscores the notion that patterns in algorithm preference demonstrate a high degree of specificity within their respective domains.

To quantify the statistical significance of this clustering effect, we designed a hypothesis test based on label permutation. After PCA projection, the coordinates of the $i$-th network are defined as $\bm{x}_i \in \mathbb{R}^3$. The domain label of the $i$-th network is denoted by $l_i$, where $l_i \in \{1, 2, \dots, 7\}$ (corresponding to the seven domains). First, we compute the observed mean intra-domain Euclidean distance as:
\begin{equation}
		 \bar{D}_{p} =  \frac{1}{u_p (u_p - 1)} \sum_{\substack{i,k \in \mathcal{N}_p \ i \neq k}} ||\bm{x}_i - \bm{x}_k||_{2},
	 \end{equation}
 where $\mathcal{N}_p$ denotes the set of networks in the $p$-th true domain, and $u_p=|\mathcal{N}_p|$.

 Subsequently, we conducted 2,000 random permutation experiments. In each experiment, we kept the number of networks in each domain constant and randomly reassigned domain labels to all networks. This process generated pseudo-label partitions denoted as $S_p, p=1, 2, \dots, 7$. We then computed the mean intra-class distance for each of these pseudo-label partitions. For the $S_p$, we can obtained that
\begin{equation}
	\bar{D}_{\text{rand}, p}=  \frac{1}{|S_p| (|S_p| - 1)} \sum_{i,k \in S_p, i \neq k} \|\bm{x}_i - \bm{x}_k\|_2.
\end{equation}
$\bar{D}_{\text{rand},p}$ represents the average intra-class Euclidean distance across the pseudo-label partitions $S_p$ after completely randomizing the domain labels. However, it is crucial to highlight that, with node labels assigned randomly, $\bar{D}_{\text{rand}, p} = \bar{D}_{\text{rand}, q}$ holds for all $p, q \in \{1, 2, \dots, 7 \}$. Moreover, this value is exactly equivalent to the expected distance between two randomly chosen nodes within the entire node set, which is expressed as 
$$\bar{D}_{\text{rand}, p} = \bar{D}_{\text{rand}} = \frac{1}{|S_\text{net}| (|S_\text{net}| - 1)} \sum_{i,k \in S_\text{net}, i \neq k} \|\bm{x}_i - \bm{x}_k\|_2, $$ where $S_\text{net}$ denotes the ensemble of all real-world networks investigated herein. Here, $S_\text{net}=740$. If the mean intra-domain Euclidean distance $\bar{D}_p$ under true labels is significantly smaller than the distribution of $\bar{D}_{\text{rand}}$ obtained from random permutations, it indicates that networks from the same domain do indeed cluster together. As shown in \autoref{fig_pca}(b), except for technological network, the mean intra-domain distances for the remaining six domains are significantly lower than the random expectation, further confirming that \textbf{domain attributes are a key factor shaping algorithmic performance patterns}.

 \subsection{Ranking stability}
 The above analysis reveals significant domain dependence in the performance of link prediction algorithms. However, one critical question that directly impacts experimental reproducibility and evaluation efficiency remains unresolved: How many networks within a specific domain are needed to obtain a stable estimate of algorithm performance? To address this issue systematically, we examine the convergence behavior of the mean algorithm ranking as the network sample size $L$ increases. For any given domain, $p$, containing $u_p$ networks, we compute the average ranking vector $\bm{r}_p$ for these $12$ link prediction algorithms.
	
\begin{figure}[htbp]
	\centering
	\centerline{\includegraphics[width=1.0\linewidth]{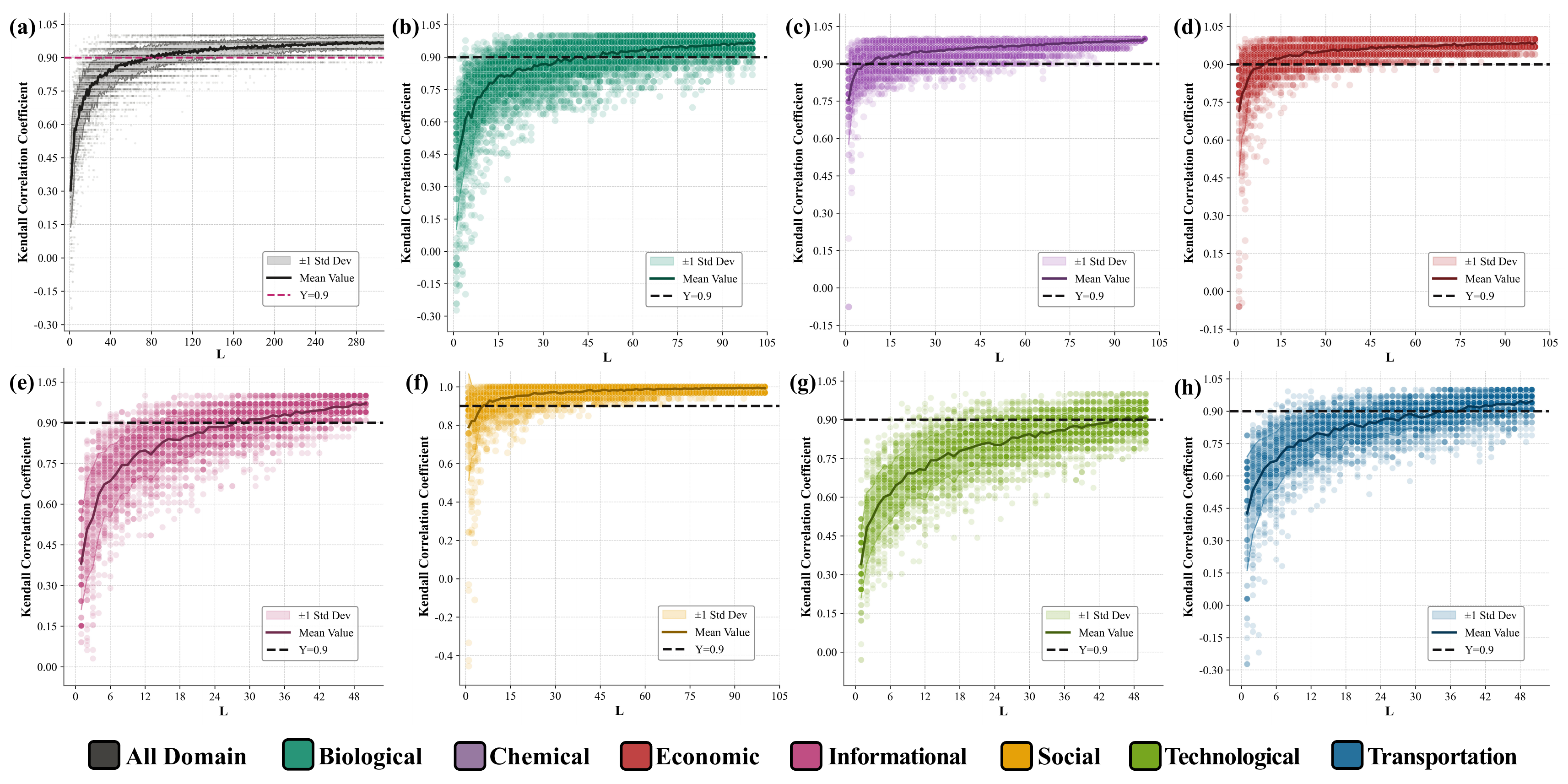}}
	\caption{{\bf  Performance and variability of link prediction algorithms $DW$ with varying sample sizes across domains.} The x-axis represents the number of randomly sampled networks $L$. The y-axis shows the ranking stability coefficient $\bar{\tau}_{L}$ of the link prediction algorithm $DW$. (a) illustrates the aggregated performance across all domains. (b)--(h) present the results for distinct domains. In each subfigure, the dashed line indicates the threshold value 0.9, and the solid line represents the mean RSC value across 100 independent trials.}
	\label{fig_size_mfi}
\end{figure}

 Subsequently, for each sample size $L$ (ranging from 1 to $u_p$, where $u_p$ is the total number of networks in domain $p$), we repeated the following process $100$ times. First, we randomly sampled $L$ networks from the domain without replacement. Then, we computed the average ranking vector $\bm{r}_L^i$ ($i=1, 2, \dots, 100$). The Kendall's $\tau$ correlation coefficient, denoted by $\tau( \bm{r}^{i}_{L}, \bm{r}_{p})$, is to be computed for all values of i ranging from $1$ to $100$, in conjunction with the reference vector. The mean of these $100$ correlation coefficients is defined as the Ranking Stability Coefficient (RSC):
\begin{equation}
	\bar{\tau}_{L}=\frac{1}{100} \sum_{i=1}^{100} \tau (\bm{r}^{i}_{L}, \bm{r}_{p}),
\end{equation}
This $\bar{\tau}_L$ value corresponds to the solid line on the vertical axis in \autoref{fig_size_mfi}.

 As demonstrated in \autoref{fig_size_mfi}, $\bar{\tau}_{L}$ rises rapidly with increasing $L$ and approaches saturation, indicating that \textbf{in most domains, only a small number of networks are needed to obtain stable algorithm ranking estimates}. For instance, if we define $\bar{\tau}_L \geq 0.90$ as a state of high consistency, then in transportation, technological, and informational networks, $\bar{\tau}_{L}$ exceeds $0.90$ when $L=10$. In contrast, biological, social, chemical, and economic networks converge more slowly, typically requiring $L>75$ to achieve comparable consistency. This phenomenon indicates high internal structural heterogeneity within these domains. Consequently, when fewer networks are used in experiments, the algorithm ranking becomes susceptible to the influence of individual networks, leading to biased algorithm ranking estimates. Notably, $\bar{\tau}_L$ stabilizes for all domains when $L \geq 90$. This finding indicates that \textbf{in practical evaluations, the uncritical pursuit of large-scale network collections is unnecessary. Rather, ensuring domain representativeness via judicious sample size control can optimize the balance between evaluation reliability and computational efficiency}. Excessively increasing the number of networks has been shown to reduce marginal gains and may lead to unnecessary computational overhead.
	
\section{Conclusion}
This study systematically examines the performance patterns of link prediction algorithms across seven application domains using 740 real-world network datasets. Analysis reveals strong domain dependency in algorithmic performance: For instance, most algorithms exhibit high and consistent AUC values in social networks. Conversely, in economic networks, despite the presence of relatively consistent algorithm rankings, the overall predictive performance is considerably suboptimal. The evidence presented herein indicates that superior algorithmic performance in one domain rarely transfers to networks of other domains, as domain attributes inherently serve as implicit yet critical structural moderators.
	
To further characterize algorithmic ranking consistency, we adopted a quantitative approach to assess the similarity of algorithm rankings both intra-domain and inter-domain. The findings indicate a high degree of intra-domain algorithmic ranking consistency ($\tau > 0.5$) in chemical, social, and economic networks, while biological and transportation networks exhibit notable intra-domain ranking instability. Of particular significance is the observation that inter-domain algorithmic ranking consistency is generally weak (average $\tau \approx 0.0991$), which is significantly lower than intra-domain levels (average $\tau \approx 0.4117$). Some domain pairs even exhibit negative correlations, indicating that current mainstream link prediction algorithms lack inter-domain transferability. This finding is further corroborated by PCA: After representing each network as its algorithm performance ranking vector across the $12$ link prediction algorithms, their low-dimensional embeddings naturally cluster by domain in PCA space, with this clustering pattern being statistically significant and non-random (except for technological networks). This finding underscores the notion that the ``algorithmic performance fingerprint'' itself contains domain-specific identity information, thereby highlighting the substantial impact of domain attributes on algorithm performance patterns.
	
Moreover, the application of the Winner Score metric to identify dominant algorithms across domains revealed the absence of a universally optimal link prediction algorithm. NMF exhibits a dominant performance in social networks, NeoGNN exhibits superior performance in economic networks, GCN exhibits optimal performance in chemical networks, and RA3 demonstrates the highest level of competitiveness in biological networks. It is noteworthy that these dominant algorithms frequently exhibit substandard or below-average performance in other domains. This observation underscores the notion that the effectiveness of an algorithm depends on the congruence between its inherent mechanisms (\textit{e.g.}, low-rank matrix modeling, high-order path propagation) and the topological characteristics of the target network. This congruence is more significant than the complexity of the model. Therefore, assuming a ``universally optimal'' algorithm divorced from specific application contexts lacks empirical basis. Finally, addressing practical concerns in experimental design, we evaluated the minimum number of networks required to obtain stable algorithm performance estimates. We defined the Ranking Stability Coefficient (RSC) to determine the number of networks required for algorithm rankings to stabilize. The results of this study demonstrated that most domains, including but not limited to transportation, informational, and technological domains, require approximately $10$ networks for algorithm rankings to stabilize (with a median RSC of $0.90$). In contrast, domains with higher structural heterogeneity, such as biological, social, and economic domains, require over $75$ networks for algorithm rankings to stabilize. The RSC saturates for all domains when the sample size is at least $90$, indicating that moderate control of the evaluation scale can balance evaluation reliability and computational efficiency while ensuring domain representativeness. This obviates the need for the blind pursuit of large-scale probe sets
	
In summary, this study establishes ``domain specificity'' as a fundamental perspective for understanding the behavior of link prediction algorithms via multidimensional empirical analysis. The findings challenge the notion of a universally optimal link prediction algorithm and demonstrate that a network's application domain is not merely a classification label for data sources, but a structural context that systematically modulates algorithmic performance patterns. This topological differentiation driven by domain differences poses fundamental challenges to the direct transfer of models trained in one domain to another. It also highlights the necessity of establishing domain-specific evaluation benchmarks and algorithm selection strategies. Consequently, this study lays a solid foundation for practical algorithm selection, benchmark construction, and evaluation practices in link prediction. This work emphasizes the necessity of evaluating the effectiveness of link prediction algorithms within their specific application contexts. Future benchmark development, algorithm comparison, and experimental design should prioritize refined research that aligns with the ``problem-data-mechanism'' framework. It is imperative to emphasize that only through this approach can the scientific rigor, reproducibility, and practical applicability of link prediction be improved.

\bibliographystyle{plain}

\newpage

\section*{Appendix: Domain-Specific Performance of Considered Algorithms}
\small
\setlength{\tabcolsep}{3pt} 
\setlength{\extrarowheight}{1pt} 

\begin{longtable}{@{}L{3cm} L{2cm} *{6}{C{1.5cm}}@{}}	
	\caption{\textbf{Performance of the twelve representative algorithms evaluated by alternative metrics.} All results are obtained by averaging over networks in each domain.}
	\label{tab:all_domains_performance} \\
	\toprule
	\small \textbf{Domain} 
	& \small \textbf{Algorithm} 
	& \small \textbf{Precision} 
	& \small \textbf{AUC-Precision} 
	& \small \textbf{AUPR} 
	& \small \textbf{AUC-mROC} 
	& \small \textbf{NDCG} 
	& \small \textbf{H-measure} \\
	\midrule
	\endfirsthead

	\caption[]{(Continued) Performance of the twelve representative algorithms evaluated by alternative metrics. All results are obtained by averaging over networks in each domain.} \\
	\toprule
	\small \textbf{Domain} 
	& \small \textbf{Algorithm} 
	& \small \textbf{Precision} 
	& \small \textbf{AUC-Precision} 
	& \small \textbf{AUPR} 
	& \small \textbf{AUC-mROC} 
	& \small \textbf{NDCG} 
	& \small \textbf{H-measure} \\
	\midrule
	\endhead

	\midrule
	\multicolumn{8}{r}{\textit{Continued on next page}} \\
	\endfoot

	\bottomrule
	\endlastfoot  
	
	\multirow{12}{3cm}{\textbf{Biological  Networks}} 
	& DW        & 0.5728    & 0.5991        & 0.6009  & 0.5819   & 0.8588  & 0.1540    \\ 
	& GAE       & 0.6202    & 0.7482        & 0.7006 & 0.7161   & 0.9093  & 0.3388    \\ 
	& GAT       & 0.6076    & 0.7002        & 0.6598 & 0.6579   & 0.8888  & 0.2900    \\ 
	& GCN       & 0.5969    & 0.6807        & 0.6583  & 0.6575   & 0.8827 & 0.2010    \\ 
	& GraphSAGE & 0.5920    & 0.7054        & 0.6598 & 0.6711   & 0.8920  & 0.2710    \\ 
	& MFI       & 0.6028    & 0.6552        & 0.6527  & 0.6143   & 0.8646  & 0.2521    \\ 
	& NeoGNN    & 0.6475    & 0.7160        & 0.7146  & 0.6769   & 0.9019  & 0.4516    \\ 
	& NMF       & 0.5884    & 0.6496        & 0.6429  & 0.6180   & 0.8678 & 0.2186    \\ 
	& PULL      & 0.6599    & 0.7901        & 0.7460  & 0.7534   & 0.9194 & 0.4360    \\ 
	& RA        & 0.3105    & 0.4515        & 0.5992  & 0.5833   & 0.8086 & 0.1402    \\ 
	& RA3       & 0.5910    & 0.7634        & 0.8061  & 0.8049   & 0.9208 & 0.3449    \\ 
	& VAGE      & 0.6526    & 0.7697        & 0.7259 & 0.7375   & 0.9194  & 0.4088    \\
	\midrule
	
	\multirow{12}{3cm}{\textbf{Chemical Networks}}   
	& DW        & 0.7616    & 0.8311        & 0.8134  & 0.8031   & 0.9354  & 0.5232    \\ 
	& GAE       & 0.4425    & 0.4185        & 0.4777  & 0.4298   & 0.7468  & 0.0536    \\ 
	& GAT       & 0.5833    & 0.6753        & 0.6550  & 0.6603   & 0.8718  & 0.2763    \\ 
	& GCN       & 0.7772    & 0.8407        & 0.8244 & 0.8112   & 0.9357  & 0.5563    \\ 
	& GraphSAGE & 0.5727    & 0.6394  & 0.6291  & 0.6262   & 0.8529  & 0.2509    \\ 
	& MFI       & 0.7699    & 0.8538      & 0.8315  & 0.8254   & 0.9406  & 0.5541    \\ 
	& NeoGNN    & 0.5000    & 0.4738      & 0.5619  & 0.4739   & 0.7744  & 0.2546    \\ 
	& NMF       & 0.6893    & 0.8190        & 0.7676  & 0.7798   & 0.9212  & 0.4047    \\ 
	& PULL      & 0.6557    & 0.7029        & 0.7104  & 0.6965   & 0.8743  & 0.4436    \\ 
	& RA        & 0.5781    & 0.7392        & 0.7723 & 0.7606   & 0.8854 & 0.3180    \\ 
	& RA3       & 0.6618    & 0.7565        & 0.7583  & 0.7446   & 0.8929  & 0.3548    \\ 
	& VAGE      & 0.4244    & 0.3990        & 0.4610  & 0.4114   & 0.7368  & 0.0277    \\
	\midrule
	
	\multirow{12}{3cm}{\textbf{Social Networks}}     
	& DW        & 0.8763 & 0.9290 & 0.9196  & 0.8993 & 0.9791  & 0.7148 \\ 
	& GAE       & 0.6592 & 0.7917 & 0.7482  & 0.7688 & 0.9505  & 0.4865 \\ 
	& GAT       & 0.6547 & 0.7758 & 0.7374 & 0.7553 & 0.9473  & 0.4840 \\ 
	& GCN       & 0.8872 & 0.9503 & 0.9355  & 0.9020 & 0.9859  & 0.7351 \\ 
	& GraphSAGE & 0.6777 & 0.7815 & 0.7424  & 0.7451 & 0.9473 & 0.4986 \\ 
	& MFI       & 0.9262 & 0.9617 & 0.9504  & 0.9376 & 0.9869 & 0.8354 \\ 
	& NeoGNN    & 0.5783 & 0.6019 & 0.6385  & 0.5545 & 0.8907 & 0.3572 \\ 
	& NMF       & 0.9212 & 0.9569 & 0.9457  & 0.9350 & 0.9848  & 0.8291 \\ 
	& PULL      & 0.7709 & 0.8754 & 0.8458  & 0.8271 & 0.9676  & 0.7025 \\ 
	& RA        & 0.9047 & 0.9606 & 0.9556  & 0.9601 & 0.9865  & 0.8338 \\ 
	& RA3       & 0.9047 & 0.9697 & 0.9561  & 0.9526 & 0.9892  & 0.7996 \\ 
	& VAGE      & 0.6714 & 0.7967 & 0.7559  & 0.7698 & 0.9521  & 0.5133 \\
	\midrule
	
	\multirow{12}{3cm}{\textbf{Economic Networks}}   
	& DW        & 0.5699 & 0.6363 & 0.6000  & 0.6055 & 0.8887  & 0.0854 \\ 
	& GAE       & 0.5154 & 0.6681 & 0.6017  & 0.6647 & 0.9000  & 0.0908 \\ 
	& GAT       & 0.4801 & 0.6238 & 0.5724  & 0.6374 & 0.8887  & 0.0771 \\ 
	& GCN       & 0.5458 & 0.5689 & 0.5697  & 0.5958 & 0.8763  & 0.0970 \\ 
	& GraphSAGE & 0.3915 & 0.4972 & 0.4867  & 0.5140 & 0.8362  & 0.0605 \\ 
	& MFI       & 0.2063 & 0.3828 & 0.4954  & 0.5624 & 0.8172  & 0.0764 \\ 
	& NeoGNN    & 0.6171 & 0.6648 & 0.6743  & 0.6336 & 0.8940  & 0.3761 \\ 
	& NMF       & 0.4408 & 0.4892 & 0.5346  & 0.5760 & 0.8589  & 0.1327 \\ 
	& PULL      & 0.5464 & 0.7131 & 0.6502  & 0.6938 & 0.9060  & 0.1886 \\ 
	& RA        & 0.0966 & 0.1276 & 0.4930  & 0.4882 & 0.7346  & 0.0489 \\ 
	& RA3       & 0.1593 & 0.3353 & 0.7408  & 0.7191 & 0.8184  & 0.0863 \\ 
	& VAGE      & 0.5498 & 0.7013 & 0.6305  & 0.6865 & 0.9112  & 0.1461 \\
	\midrule
	
	\multirow{12}{3cm}{\textbf{Transportation Networks}} 
	& DW        & 0.6205 & 0.7207 & 0.6795  & 0.6925 & 0.9144 & 0.1833 \\ 
	& GAE       & 0.5430 & 0.5797 & 0.5839  & 0.5810 & 0.8672 & 0.1883 \\ 
	& GAT       & 0.6011 & 0.6932 & 0.6540  & 0.6661 & 0.9077 & 0.2878 \\ 
	& GCN       & 0.6400 & 0.7264 & 0.6998  & 0.6992 & 0.9120 & 0.2420 \\ 
	& GraphSAGE & 0.5665 & 0.6393 & 0.6112  & 0.6276 & 0.8878  & 0.2029 \\ 
	& MFI       & 0.5918 & 0.7520 & 0.7245  & 0.7305 & 0.9086 & 0.3149 \\ 
	& NeoGNN    & 0.5516 & 0.5868 & 0.6051  & 0.5601 & 0.8628 & 0.2542 \\ 
	& NMF       & 0.6140 & 0.7427 & 0.7096  & 0.7104 & 0.9127 & 0.2808 \\ 
	& PULL      & 0.6364 & 0.7121 & 0.6840  & 0.6855 & 0.9029 & 0.3710 \\ 
	& RA        & 0.2222 & 0.3830 & 0.7145  & 0.6866 & 0.8174 & 0.1168 \\ 
	& RA3       & 0.3170 & 0.5315 & 0.7611  & 0.7410 & 0.8572  & 0.1740 \\ 
	& VAGE      & 0.5353 & 0.5776 & 0.5799  & 0.5842 & 0.8666  & 0.1795 \\
	\midrule
	
	\multirow{12}{3cm}{\textbf{Technological Networks}} 
	& DW        & 0.5708 & 0.6101 & 0.5965  & 0.5982 & 0.8835 & 0.1101 \\ 
	& GAE       & 0.6120 & 0.7432 & 0.6922  & 0.7254 & 0.9213  & 0.3122 \\ 
	& GAT       & 0.6016 & 0.7288 & 0.6722  & 0.7007 & 0.9179 & 0.2755 \\ 
	& GCN       & 0.6246 & 0.7361 & 0.6914  & 0.7060 & 0.9172  & 0.2073 \\ 
	& GraphSAGE & 0.5818 & 0.6977 & 0.6478 & 0.6779 & 0.9060 & 0.2267 \\ 
	& MFI       & 0.6422 & 0.7456 & 0.6976 & 0.6765 & 0.9107 & 0.2463 \\ 
	& NeoGNN    & 0.6061 & 0.6553 & 0.6699  & 0.6134 & 0.8896  & 0.3839 \\ 
	& NMF       & 0.6002 & 0.7267 & 0.6788  & 0.6716 & 0.9083  & 0.2114 \\ 
	& PULL      & 0.6585 & 0.7825 & 0.7341 & 0.7527 & 0.9264 & 0.4178 \\ 
	& RA        & 0.2544 & 0.4227 & 0.6122  & 0.6329 & 0.8220  & 0.1292 \\ 
	& RA3       & 0.4005 & 0.6141 & 0.7172  & 0.7432 & 0.8818  & 0.2119 \\ 
	& VAGE      & 0.6170 & 0.7502 & 0.6993  & 0.7305 & 0.9238 & 0.3325 \\
	\midrule
	
	\multirow{12}{3cm}{\textbf{Informational\\Networks}} 
	& DW        & 0.6312 & 0.7016 & 0.6784  & 0.6680 & 0.9112  & 0.2211 \\ 
	& GAE       & 0.6869 & 0.8436 & 0.7864  & 0.8079 & 0.9512  & 0.4956 \\ 
	& GAT       & 0.6552 & 0.7917 & 0.7393 & 0.7450 & 0.9361 & 0.4238 \\ 
	& GCN       & 0.6860 & 0.8125 & 0.7709 & 0.7792 & 0.9393  & 0.3544 \\ 
	& GraphSAGE & 0.6402 & 0.7786 & 0.7253  & 0.7444 & 0.9300  & 0.3838 \\ 
	& MFI       & 0.6607 & 0.7917 & 0.7828  & 0.7392 & 0.9274  & 0.3846 \\ 
	& NeoGNN    & 0.6569 & 0.7344 & 0.7192  & 0.6714 & 0.9159 & 0.4591 \\ 
	& NMF       & 0.6367 & 0.7630 & 0.7245 & 0.7216 & 0.9292  & 0.2982 \\ 
	& PULL      & 0.7250 & 0.8644 & 0.8177 & 0.8267 & 0.9533  & 0.5791 \\ 
	& RA        & 0.5367 & 0.7559 & 0.8195  & 0.8182 & 0.9211 & 0.3487 \\ 
	& RA3       & 0.6263 & 0.8106 & 0.8395  & 0.8364 & 0.9343  & 0.4049 \\ 
	& VAGE      & 0.6999 & 0.8505 & 0.7963  & 0.8122 & 0.9547 & 0.5253 \\
	
\end{longtable}
\end{document}